\documentstyle[preprint,revtex]{aps}
\begin{document}
\draft
\preprint{}
\begin{title}
Geometric and Diffractive Orbits in the Scattering from Confocal 

Hyperbolae
\end{title}
\author{N. D. Whelan}
\begin{instit}
Centre for Chaos and Turbulence Studies

Niels Bohr Institute, Blegdamsvej 17, DK-2100, Copenhagen \O, Denmark
\end{instit}
\date{\today}
\begin{abstract}

We study the scattering resonances between two confocal hyperbolae and show 
that the spectrum is dominated by the effect of a single periodic
orbit. There are two distinct cases depending on whether the orbit is 
geometric or diffractive. A generalization of periodic orbit theory allows us 
to incorporate the second possibility. In both cases we also perform a WKB 
analysis. Although it is found that the semiclassical approximations work 
best for resonances
with large energies and narrow widths, there is reasonable agreement
even for resonances with large widths - unlike the two disk scatterer. 
We also find agreement with the next order correction to periodic orbit 
theory.

\end{abstract}
\pacs{PACS numbers: 03.20 03.65.Sq}

\narrowtext
 
In recent years there has been growing interest in understanding the extent to
which knowledge of classical mechanics can be used used to understand 
quantum systems \cite{gutz}.  For sufficiently hyperbolic systems, periodic
orbits provide an efficient means of calculating spectra semi-classically. One
class of problem which has been fruitfully studied is the n-disk scatterer in
two dimensions \cite{disk1,disk2}.  It has been shown that in some situations
periodic orbit theory gives the energies and widths of the scattering
resonances to great accuracy - typically several decimal places.

However, this accuracy is usually only for the leading family of resonances.
There exist
other resonances with larger widths deeper in the complex momentum plane.
These are badly approximated by periodic orbit theory because
it fails to consider classical paths which, although not trajectories,
satisfy the stationary phase  condition. Recent work \cite{creep1,creep2}
has shown that inclusion of diffractive paths - so-called creeping orbits - 
recovers
the qualitative features of the exact spectrum, including the lower order
families.  However, the quantitative agreement is not as good as for the 
leading family of resonances which are unaffected by creeping.

It is useful to consider a system which has no creeping so as to study
the full spectrum of resonances.  One such system is a pair of 
confocal hyperbolae.  Like the two-disk
scatterer there is only one periodic orbit, which is unstable. However,
unlike the two disk scatterer, there are
no creeping orbits. Since the system is separable we can apply WKB
techniques as well as periodic orbit theory but this is more difficult 
and is not as intuitive. In addition, we can evaluate
the resonances for the special case where one or both hyperbolae are close to
half planes. Then we must consider edge diffraction and
this is included in the periodic orbit analysis.

We want to solve the Schr\"{o}dinger equation in a domain between two confocal
hyperbolae, as shown for example in Figs.~1 and 2. This is simply the
Helmholtz equation $(\nabla^2+k^2)\Psi = 0$ with the boundary
conditions that on the two hyperbolae $\Psi$ vanishes and that in the region
between them $\Psi$ approaches
$f(\phi)\exp(ikr)/\sqrt{r}$ for large $r$ (where $r$ and $\phi$ are polar
coordinates.)  We will work in hyperbolic-elliptic coordinates,
$\mu$ and $\theta$, defined by $x = a\cosh{\mu}\cos{\theta}$ and
$y = a\sinh{\mu}\sin{\theta}$ \cite{eigen}. Curves of constant
$\mu$ are ellipses and curves of constant $\theta$ are hyperbolae. In both
cases the foci are at $y=0$ and $x=\pm a$. We take the coordinates to be in 
the ranges $0\leq\theta\leq\pi$ and $-\infty<\mu<\infty$ and label the right 
and left hyperbolae respectively by $\theta_1$ and $\theta_2$ so that
$\theta_2>\theta_1$. We define length units such that the closest distance 
between the hyperbolae 
is 1. It follows that $a=1/(\cos{\theta_1}-\cos{\theta_2})$. This can be
generalized to any spacing $L$ by substituting $kL$ for $k$ in what follows.

Using the separation $\Psi(\vec{r})=M(\mu)\Theta(\theta)$ and the expression
$\nabla^2=\Bigl(\frac{\partial^2}{\partial\mu^2}+\frac{\partial^2}
{\partial\theta^2}\Bigr)/a^2(\cosh^2\mu-\cos^2\theta)$, we obtain the Mathieu 
equations
\begin{mathletters}  \label{eq:mathieu}
\begin{equation} \label{eq:matha}
\frac{d^2\Theta}{d\theta^2} + a^2(b^2-k^2\cos^2\theta)\Theta = 0
\end{equation}
\begin{equation} \label{eq:mathb}
\frac{d^2M}{d\mu^2} + a^2(-b^2+k^2\cosh^2\mu)M = 0,
\end{equation}
\end{mathletters}
where $a^2b^2$ is a separation constant.
Although we can work with these equations, it is convenient to express equation
(\ref{eq:mathb}) as a one dimensional potential problem.  This is done
by a change of coordinates $\zeta=a\sinh{\mu}$ and
$M(\mu)=Z(\zeta)/(\zeta^2+a^2)^{1/4}$ yielding
\begin{equation}  \label{eq:radmod}
\frac{d^2 Z}{d\zeta^2} + \bigl(k^2 - 2V(\zeta)\bigr)Z = 0,
\end{equation}
with an effective potential
\begin{equation} \label{eq:pot}
V(\zeta) = \frac{1}{2}\left(\frac{b^2a^2+1/2}{\zeta^2+a^2} -
\frac{3\zeta^2}{4(\zeta^2+a^2)^2}\right).
\end{equation}
The asymptotic boundary condition is $Z(\zeta)\sim\exp(ik\zeta)$. There is a
$y\rightarrow -y$ symmetry so there are two symmetry classes;
$Z'(0)=0$ for even states and $Z(0)=0$ for odd states. The effective potential
is repulsive so there are no bound states, only resonances.

Resonances occur for negative imaginary $k$ which 
means that asymptotically the wave function is increasing 
exponentially \cite{landau} as ${\cal O}(\exp(k_i\zeta))$
where $k=k_r-ik_i$. Such wavefunctions are called 
Siegert or Gamow states and their asymptotic nature is a reflection of the
time-dependent decay. To work with such wave functions numerically, one
must complexify the coordinate $\zeta=|\zeta|\exp(i\alpha)$ \cite{comcor}.
Asymptotically the wavefunctions are decaying exponentially 
as a function of $|\zeta|$ if $\alpha>-\arg(k)$.

We numerically solve the simultaneous equations for $k$ and $b$ by shooting
\cite{num_rec}. The exact spectra for two choices of parameters are 
shown in Figs.~1 and 2 as open diamonds. Selected data are also listed in 
Tables I and II. In Fig.~1, $\theta_1=0.4$ and
$\theta_2=0.9$ which is a smooth system since the radii of curvature of the
hyperbolae are comparable to the inter-hyperbola spacing of 1. In Fig.~2,
$\theta_1=0$ and $\theta_2=2.0$ which is sharp since the right hyperbola is a
half plane and its radius of curvature is zero.  
As will be discussed, the theoretical understanding of the
two cases is quite different. For each value of 
$k$ there is one corresponding value of $b$.  The spectra of $k$ and $b$
are similar in both cases; their real components are very close and their
imaginary components differ approximately by a constant.

Periodic orbit theory works by approximating the trace of the Green function,
the poles of which are bound states or scattering resonances \cite{gutz}. 
This has been worked out for the two disk problem \cite{creep1} in terms of the
single periodic orbit in the system. The result, adapted to this system and
including both symmetry classes, is
\begin{equation} \label{eq:perorbt}
k_{nj} = (n+1)\pi - i\frac{2j+1}{4}\log\Lambda + \cdots
\end{equation}
where $n,j=0,1,\cdots$ and $\Lambda = \tan^2(\theta_2/2)/\tan^2(\theta_1/2)$  
is the stability for a complete traversal \cite{whelan}.
The $y\rightarrow -y$ parity is $(-1)^j$.

Equations (\ref{eq:matha}) and (\ref{eq:radmod}) can also be solved using WKB
theory.  The details will be presented elsewhere \cite{whelan}. We begin with
the expansion
\begin{eqnarray}
k & = & k_r - ik_i + {\cal O}(k^{-1})  \label{eq:expansh}\\
b & = & b_r - ib_i + {\cal O}(b^{-1}). \nonumber
\end{eqnarray}
$b$ is the eigenvalue of the equation $({\cal L} + b^2)\Psi=0$ where ${\cal L}
=(\cosh^2\mu\frac{\partial^2}{\partial\theta^2} +
\cos^2\theta\frac{\partial^2}{\partial\mu^2})/(\cosh^2\mu-\cos^2\theta)$ as can
be seen from equation (1).
In the neighbourhood of the periodic orbit and to leading order,
the equation for $b$ is the same as the equation for $k$ so that $b_r=k_r$.

The WKB approximation for (\ref{eq:matha}) is \cite{bermount}
\begin{equation} \label{eq:wkb1}
(n+1)\pi = a\int_{\theta_1}^{\theta_2}d\theta\sqrt{b^2-k^2\cos^2\theta}
\end{equation}
This equation can be expressed in terms of incomplete elliptic
integrals \cite{absteg}.  We expand the integrand of equation (\ref{eq:wkb1})
in powers of 
$(b^2-k^2)/(k^2\sin^2\theta)$ and keep the first two terms.  This gives
$k_r=(n+1)\pi$ and $b_i=fk_i$ where $f=(1-2/a\log\Lambda)$. The
imaginary components of $k$ and $b$ are related by a factor $f$ which depends 
only on the geometry and is the same for all resonances.

The WKB approximation for the resonances of equation (\ref{eq:radmod}) is 
\cite{bermount}
\begin{equation} \label{eq:wkb2}
(2j+1)\pi=2a\int_{z_-}^{z_+}dz\sqrt{k^2-2V(z)}
\end{equation}
where $V(z)$ is shown in equation (\ref{eq:pot}). This formula comes from 
identifying the
resonances as the poles of the transmission coefficient. $z_{\pm}$ are complex
turning points which are solutions of $V(z_{\pm})=k^2/2$. We 
arrive at a solution in which $k$ is as in equation (\ref{eq:perorbt}) and
\begin{equation} \label{eq:wkbres}
b_{nj} = (n+1)\pi - if\frac{2j+1}{4}\log\Lambda + \cdots
\end{equation}

These results are consistent with periodic orbit theory.  However the
calculation of $b$ is new. The imaginary component of $b$ is less negative
than that of $k$ while the real components are equal
to leading order.  The semiclassical results are shown in Fig.~1 as crosses
and are also listed in Table I.
The worsening of the agreement as we go down in the $k$ plane is 
expected \cite{voros} and is consistent with the expansion (\ref{eq:expansh}).

It is interesting to study the higher order terms in the expansion of the
resonances.  The first two corrections to the leading family of resonances 
yield
\cite{vattay,gasal} 
\begin{equation} \label{eq:corr}
k_n = (n+1)\pi - \frac{c_r}{(n+1)\pi}
-i\left(\frac{\log{\Lambda}}{4}-\frac{c_i}{(n+1)^2\pi^2}\right),
\end{equation}
where $c_r$ and $c_i$ are constants. For the symmetric case
$\theta_2=\pi-\theta_1$, we have \cite{vattay} $c_r=\cot^2\theta_1$. 
For the example $\theta_1=1.0$ we fit the 200'th resonance to equation 
(\ref{eq:corr}) to extract $c_r=0.4123$ in agreement
with the expected result.  We also find $c_i=0.593$ but there is as yet no
analytical result. The two constants have been worked out for the 
two disk problem \cite{gasal} where it is found that $c_i$ is positive.
This means that the resonance lifetimes are longer than the first order 
approximation, as is also found here.
Since periodic orbit theory is an asymptotic expansion in $\hbar$ 
\cite{berry} we expect that the higher order corrections will
eventually blow up for any resonance $n$.
The results for the scaling with $j$ are ambiguous so 
we leave this for a later publication.

Note that some of the resonances are not in the physical 
region $k_r>0$ but instead have $k_r<0$. 
(Because the dimension is even, the
pole structure is not symmetric with respect to reflections through the
imaginary $k$ axis \cite{wirzba}.) For
example, the state with indices $n=0$ and $j=4$ does not exist as a physical
resonance although there is a semiclassical prediction. More resonances lie in
the unphysical region for large $j$ and for small $\theta_1$.

For the configuration of Fig.~2 we need a different analysis since equation 
(\ref{eq:perorbt}) predicts infinite widths.
The previous results relied on an expansion in powers of $1/\sin^2\theta$ which
breaks down if $\theta_1\ll 1$ or $\pi-\theta_2\ll 1$. 
In that case we can approximate the hyperbola by a wedge so that 
a qualitatively new phenomenon is responsible for the resonances.
There is a class of orbits which are not classical trajectories but
rather are diffractive \cite{keller} as shown in Fig.~2.
The ray starting at $P$ illuminates the vertex which then acts as a point 
source for an outgoing circular wave which illuminates $P'$. The Green 
function for this process \cite{keller,creep2} is
\begin{equation} \label{eq:green}
G(P',P,k) = -\frac{d\exp(i\left(kr'+\pi/4)\right)}{\sqrt{2\pi kr'}}G(V,P,k)
\end{equation}
where $G(V,P,k)$ is the free space Green function which connects the point $P$
to the vertex $V$ and can be approximated by the standard Van Vleck formula.
The geometric factor $d$ depends on the incoming and outgoing
angles and on the wedge angle. 

As with geometric orbits \cite{gutz} taking the trace of $G$ selects the
closed orbits.  These are all repetitions of the primitive orbit which 
diffracts from the wedge and bounces back off the wall, with a sign change. 
When the left wall is straight 
the primitive orbit contributes to the trace an amount $-ig_0(k)/4$, where
\begin{equation} \label{eq:prime}
g_0(k) = d\exp(i\left(2k+\pi/4)\right)/\sqrt{4\pi k}.
\end{equation}
For normal incidence
$d=\frac{\cot(\pi/2\gamma)}{1-\sec^2(\pi/2\gamma)/4}/\gamma$ \cite{whelan} with
$\gamma=2(1-\theta_1/\pi)$. Note that $d=1$ for $\theta_1=0$.
We must also include the focusing properties of the left wall.
The theory of reference \cite{keller} relies on a picture of cones such that 
amplitudes
decrease with the square root of the cone widths. For a straight wall,
a cone leaving the vertex with an opening angle $\phi$ returns with a width
$2\phi$.  For a curved wall, the cone returns with a width
$2\phi(1+1/R)$ where $R=-\sin\theta_2\tan\theta_2/(1-cos\theta_2)$
is the radius of curvature of the left wall. Therefore, the amplitude is
divided by a factor of $\sqrt{1+1/R}$ and the denominator of equation
(\ref{eq:prime}) is replaced by $\sqrt{4\pi k(1+1/R)}$. This factor can also 
be derived from the integration perpendicular to the diffractive orbit when 
the trace is evaluated but the geometric argument above is clearer.

Multiple diffractive terms in the trace are just powers of the primitive 
orbit \cite{creep2} so
\begin{equation} \label{eq:trace}
\mbox{Tr}G(k)  =  -\frac{i}{4}\sum_{r=1}^{\infty}g_0^r(k)
\end{equation}
\begin{displaymath}
= -\frac{i/4}{\exp\{ -i(2k+\pi/4)+\log\frac{1}{d}\sqrt{4\pi k(1+1/R)}\}-1}
\end{displaymath}
The poles are given to leading order by
\begin{equation} \label{eq:diffres}
k_n=(n+\frac{7}{8})\pi - \frac{i}{4}\log\frac{4\pi^2(n+7/8)(1+1/R)}{d^2}
\end{equation}
The factor of $7/8$ comes from the $\pi/4$ phase shift in equation 
(\ref{eq:green}) and is an important qualitative distinction from the geometric
theory. A simple extension to the case where both hyperbolae are sharp yields
\begin{equation} \label{eq:diff2res}
k_n=(n+\frac{3}{4})\pi - \frac{i}{2}\log\frac{2\pi^2(n+3/4)}{d_1d_2}.
\end{equation}

A WKB analysis is possible for the sharp geometry.  For simplicity, we just
discuss the result for $\theta_1=0$. Then the integral in equation
(\ref{eq:wkb1}) can be expressed as a complete elliptic integral of the
second kind \cite{absteg} and expanded in powers and logarithms of 
$(b^2-k^2)/b^2$. Again we use the expansion (\ref{eq:expansh}) and argue that 
$b_r=k_r$. The WKB analysis of equation (\ref{eq:wkb2}) is the same as before.
The result of these calculations is that $k$ is given as in equation 
(\ref{eq:diffres}) and $b_n=k_n+i/2a$.  As with the geometric case, 
the imaginary
component of $b$ is less negative than the imaginary component of $k$.

The results of the semiclassical analysis for the diffractive case are shown in
Fig.~2 and in Table~II. We find good agreement with the exact results
for both $k$ and $b$. As in the geometric case, the exact lifetimes are 
longer than the lowest order semiclassical prediction. The theory only 
predicts one family of resonances. This is consistent with the numerics where
we see the that widths of the next family increase logarithmically as
$\theta_1\rightarrow 0$.

There are two unexplained features of the spectrum for $\theta_1$ small but
nonzero.  One is the crossover to the geometric result near
$|k|\sim 1/(\pi\theta_1^{2})$ (by the criterion that $4\pi|k|\sim\Lambda$).
The other is the positions of the lower families. These might be 
explained by including both geometric and diffractive orbits in 
the Green function.  These calculations should be possible for the WKB analysis
as well. Another interesting issue is the higher order corrections which 
are calculable using either periodic orbit theory or WKB.
We can also use periodic orbit theory to approximate the resonance
spectrum in a system comprised of wedges of finite angles.
These issues will be addressed in future publications.

I would like to thank G\'{a}bor Vattay for suggesting this project 
and for showing me the results of his unpublished calculations. I benefitted
tremendously from discussions with him and with Stephen Creagh and Andreas 
Wirzba. I would also like to thank
Charles Clark for telling me about the complex coordinate rotation method.
This work was supported under the EU Human Capital and Mobility Programme.

\figure{The top box is the configuration space for $\theta_1=0.4$ and 
$\theta_2=0.9$. The sole periodic orbit is shown as a dotted line.  
The middle and bottom boxes show the spectra of $k$ and $b$ respectively,
plotted in the complex plane. The exact quantum resonances are represented
as open diamonds and the semiclassical predictions are represented as crosses.}

\figure{The same as Fig.~1 for $\theta_1=0$ and $\theta_2=2.0$. The
dotted line in the top box is the diffractive periodic orbit.  The crooked
dashed line is a typical diffractive path connecting the points $P$ 
and $P'$. Its Green function is given by equation (\ref{eq:green}).}

\begin{table}
\caption{The exact and semiclassical predictions of $k$ for a selection
of resonances and for the geometry of Fig.~1.}
\begin{tabular}{cccc}
$n$ & $j$ & $k$ & $k^{sc}$\\
\hline
 0 &  0 &  3.0587 -$i$0.3930 &  3.1416 -$i$0.4342 \\
 2 &  0 &  9.3881 -$i$0.4258 &  9.4248 -$i$0.4342 \\
10 &  0 & 34.5467 -$i$0.4334 & 34.5575 -$i$0.4342 \\
20 &  0 & 65.9677 -$i$0.4340 & 65.9734 -$i$0.4342 \\
 6 &  0 & 21.9714 -$i$0.4326 & 21.9911 -$i$0.4342 \\
 6 &  6 & 21.2799 -$i$5.5685 & 21.9911 -$i$5.6443 \\
 6 & 12 & 19.4438 -$i$10.4733 & 21.9911 -$i$10.8544 
\end{tabular}
\end{table}

\begin{table}
\caption{As in Table I for the geometry of Fig.~2.}
\begin{tabular}{cccc}
$n$ & $j$ & $k$ & $k^{sc}$\\
\hline
 0 &  0 &   2.6537 -$i$1.0060 &  2.7489 -$i$1.0201\\
 2 &  0 &   8.9913 -$i$1.3160 &  9.0321 -$i$1.3175\\
10 &  0 &  34.1514 -$i$1.6501 & 34.1648 -$i$1.6501\\
20 &  0 &  65.5731 -$i$1.8131 & 65.5807 -$i$1.8131
\end{tabular}
\end{table}

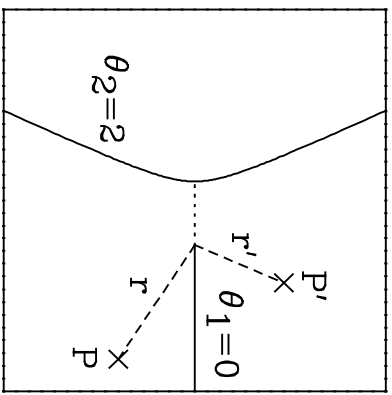
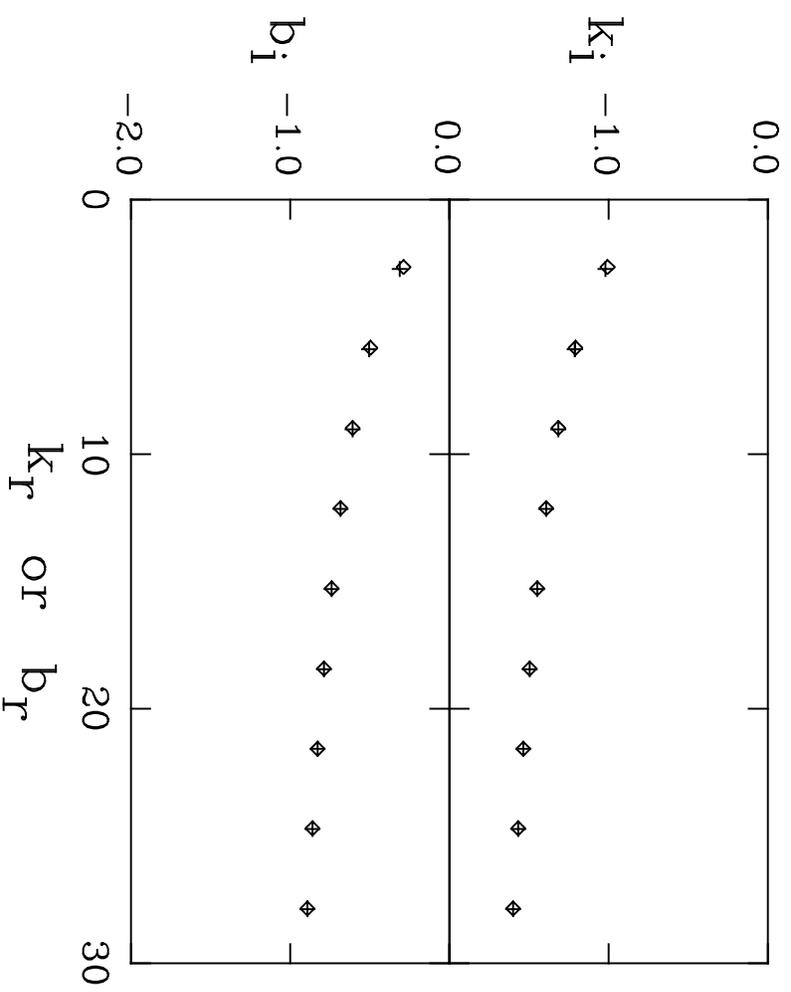
\end{document}